\documentclass[%
aps,prl,superscriptaddress,twocolumn,showpacs,preprintnumbers,amsmath,amssymb
]{revtex4-2}

\usepackage{graphicx}
\usepackage{color}
\usepackage{bm}
\usepackage{hyperref}
\usepackage{caption}
\usepackage{mathcomp}
\usepackage{ragged2e}
\usepackage{float}
\usepackage{booktabs} 

\makeatletter
\renewcommand{\@makecaption}[2]{%
  \vskip\abovecaptionskip
  \centering
  \begin{minipage}{\columnwidth} 
  \justifying
  \textbf{#1} #2 
  \end{minipage}%
  \vskip\belowcaptionskip}
\makeatother

\makeatletter
\renewcommand{\fnum@figure}{\textbf{\figurename~\thefigure $|$}} 
\makeatother

\newcommand{\extendedcaption}[1]{%
    \renewcommand{\figurename}{Extended Data Fig.} 
    \caption{#1} 
    \renewcommand{\figurename}{Fig.} 
}

\newcommand{\equalcontribution}{\textsuperscript{\textdagger}} 

\begin{document}


\title{Spin phase detection by spin current in a chiral helimagnet}

\author{Nan~Jiang\equalcontribution}
\email{nan.jiang@phys.sci.osaka-u.ac.jp}
\thanks{\equalcontribution~These authors contributed equally to this work.}
\affiliation{Department of Physics, Graduate School of Science, Osaka University, Toyonaka, Osaka 560-0043, Japan}
\affiliation{Center for Spintronics Research Network, Osaka University, Toyonaka 560-8531, Japan}
\affiliation{Institute for Open and Transdisciplinary Research Initiatives, Osaka University, Suita 565-0871,Japan}
\author{Shota~Suzuki\equalcontribution}
\affiliation{Department of Physics, Graduate School of Science, Osaka University, Toyonaka, Osaka 560-0043, Japan}
\author{Issei~Sasaki\equalcontribution}
\affiliation{Department of Physics, Graduate School of Science, Osaka University, Toyonaka, Osaka 560-0043, Japan}
\author{Kazuki~Yamada}
\affiliation{Department of Physics, Graduate School of Science, Osaka University, Toyonaka, Osaka 560-0043, Japan}
\author{Ryoma~Kawahara}
\affiliation{Department of Physics, Graduate School of Science, Osaka University, Toyonaka, Osaka 560-0043, Japan}
\author{Shintaro~Takada}
\affiliation{Department of Physics, Graduate School of Science, Osaka University, Toyonaka, Osaka 560-0043, Japan}
\affiliation{Institute for Open and Transdisciplinary Research Initiatives, Osaka University, Suita 565-0871,Japan}
\affiliation{Center for Quantum Information ad Quantum Biology, Osaka University, Toyonaka 560-0043, Japan}
\author{Yusuke~Shimamoto}
\affiliation{Department of Physics and Electronics, Osaka Metropolitan University, Sakai, Osaka 599-8531, Japan}
\author{Hiroki~Shoji}
\affiliation{Department of Physics and Electronics, Osaka Metropolitan University, Sakai, Osaka 599-8531, Japan}
\author{Yusuke~Kousaka}
\affiliation{Department of Physics and Electronics, Osaka Metropolitan University, Sakai, Osaka 599-8531, Japan}
\author{Jun-ichiro~Ohe}
\affiliation{Department of Physics, Toho University, Funabashi, Chiba 274-8510, Japan}
\author{Yoshihiko~Togawa}
\affiliation{Department of Physics and Electronics, Osaka Metropolitan University, Sakai, Osaka 599-8531, Japan}
\author{Yasuhiro~Niimi}
\affiliation{Department of Physics, Graduate School of Science, Osaka University, Toyonaka, Osaka 560-0043, Japan}
\affiliation{Center for Spintronics Research Network, Osaka University, Toyonaka 560-8531, Japan}
\affiliation{Institute for Open and Transdisciplinary Research Initiatives, Osaka University, Suita 565-0871,Japan}

\date{\today}

\begin{abstract}
Helimagnets, characterized by a helical arrangement of magnetic moments, possess unique internal degrees of freedom, including the spin phase, defined by the phase of the helical magnetic structure. Electrical detection of the spin phase is essential for both practical applications and fundamental research in helimagnets.
Here, we demonstrate the electrical detection of the spin phase in a van der Waals nanoscale chiral helimagnet CrNb$_3$S$_6$ using nonlocal spin valve measurements. Due to the short spin diffusion length in CrNb$_3$S$_6$ ($\sim5$~nm), the surface magnetic moment direction, which corresponds to the spin phase, can be detected via spin currents. The experimentally observed magnetic field dependence of the nonlocal spin valve signal is consistent with that of the surface magnetic moment in the helical magnetic structure, as supported by micromagnetic simulations.
Our results establish spin currents as a powerful tool for detecting the spin phase in helimagnets, opening avenues for utilizing the spin phase as a novel internal degree of freedom in nanoscale spintronic devices.
\end{abstract}

\maketitle


Phase is a fundamental concept in condensed matter physics, underlying phenomena such as charge density waves, Wigner crystals, Josephson junctions, and qubits. In magnetism, helimagnets~\cite{Yoshimori,Kaplan,VILLAIN1959303,NAGAMIYA1968305,DZYALOSHINSKY1958241,dzyaloshinskii1964theory,dzyaloshinskii1965theory,moriya1960anisotropic}, with their helical spin arrangements, provide a unique platform where phase plays a central role. Unlike ferromagnets and antiferromagnets, helimagnets have two internal degrees of freedom: helicity and spin phase. Helicity indicates the rotational handedness of the spin texture (right- or left-handed), while the spin phase (see Fig.~1a) specifies the position within the helical cycle. This duality gives helimagnets richer tunability, enabling novel spintronic applications.

In centrosymmetric helimagnets~\cite{Yoshimori,Kaplan,VILLAIN1959303,NAGAMIYA1968305}, helicity can be controlled by electric currents under a magnetic field~\cite{MnP,MnPtheory,MnAu2spin} and electrically detected through nonreciprocal transport~\cite{MnP,MnAu2spin} or current-induced spin accumulation~\cite{MnAu2spin}. In contrast, in noncentrosymmetric helimagnets~\cite{DZYALOSHINSKY1958241,dzyaloshinskii1964theory,dzyaloshinskii1965theory,moriya1960anisotropic}, such as chiral helimagnets, the helicity is fixed by the crystal structure and cannot be externally manipulated, although it remains electrically detectable~\cite{Yokouchi2017,CNS2w,CNSCISS}.
This makes the spin phase the primary internal degree of freedom. 

In nanoscale chiral helimagnets, the spin phase directly corresponds to the surface magnetic moment orientation, as illustrated in Fig.~1a. While surface-sensitive techniques such as x-ray magnetic circular dichroism~\cite{XMCD1,XMCD2,XMCD3,XMCD4} and spin-polarized scanning tunneling microscopy~\cite{STM1,STM2,STM3} can probe surface magnetization, their reliance on ultra-high vacuum or synchrotron radiation severely limits practical applications. 
Conventional electrical detection of magnetization, such as anomalous Hall effect measurements~\cite{nagaosa2010anomalous}, can detect only the net magnetization averaged over the conducting region, making it impossible to isolate surface contributions. 
Thus, an electrical method to detect the spin phase is crucial for practical devices.

Spin currents offer a promising alternative for detecting spin states in magnetic materials~\cite{han2020spin}. In these materials, the spin diffusion length, that is the characteristic distance over which a spin current decays, is typically on the order of a few nanometers. As a result, injected spin currents mainly interacts with surface magnetic moments. This inherent surface sensitivity makes spin currents ideal for probing the spin phase in nanoscale helimagnets. Furthermore, nonlocal spin valve (NLSV) devices utilizing spin currents~\cite{NL0,NL1,NL2,NL3,NL4,Niimi_2015} provide a sensitive platform for detecting magnetization directions and spin fluctuations~\cite{PdNi,FePt,NiCu,SHE_Cr,SHE_Nagaosa}, enabling novel spintronic applications.

Here, we demonstrate electrical spin phase detection in a nanoscale helimagnet using spin currents. As a simple yet effective platform, we focus on a van der Waals (vdW) chiral helimagnet CrNb$_3$S$_6$, which possesses only one helicity. Using a NLSV device with CrNb$_3$S$_6$, we successfully detect the spin phase in nanoscale chiral helimagnets, supported by micromagnetic simulations. Additionally, inverse spin Hall effect (ISHE) measurements reveal spin fluctuations in chiral helimagnets, opening a route to studying chiral spin dynamics. Our results establish spin currents as a powerful probe for both electrical spin phase detection and spin fluctuation analysis, highlighting the spin phase as a key degree of freedom in nanoscale spintronics.

\section{spin current device with a chiral helimagnet}

CrNb$_3$S$_6$~\cite{CNS0,CNS1,CNS2,CNS3,CNS4,CNS5,CNS6,CNS7,CNS8,CNS9,CNS10,CNS11,CNS12,Kousaka2022} is a chiral helimagnet with a noncentrosymmetric crystal structure belonging to the space group $P$6$_3$22. It consists of hexagonal NbS$_2$ planes with Cr atoms intercalated at one-third of the vdW gaps, as illustrated in Fig.~1b. Due to the Dzyaloshinskii-Moriya (DM) interaction, CrNb$_3$S$_6$ exhibits a helical magnetic phase below $T_{\rm C} \sim 130$~K~\cite{Kousaka2022}. The helical axis runs along the $c$-axis, forming a chiral soliton lattice under an in-plane magnetic field~\cite{CNS7}. CrNb$_3$S$_6$ can be cleaved along the helical axis using mechanical exfoliation, achieving thicknesses of several tens of nanometers~\cite{CNS_1pich}.

To detect the spin phase via spin currents, we fabricated a lateral spin valve with insertion of a CrNb$_3$S$_6$ flake between two permalloy (Py) wires, all connected by a Cu wire (Fig.~1c). 
The thickness $t$ of the CrNb$_3$S$_6$ thin flakes used in this work can be classified into two types: one with a thickness approximately equal to the length of one helical period of CrNb$_3$S$_6$ ($L_{0}$~(= 48~nm)) under zero magnetic field~\cite{CNS4,CNS7}, and the other with a thickness approximately equal to 1.5$L_{0}$. These values are confirmed via atomic force microscopy and magnetoresistance measurements. Figures~1d and 1e show the magnetoresistances of CrNb$_3$S$_6$ thin flakes with $t = 43$~nm ($\approx L_{0}$) and $t = 72$~nm ($\approx 1.5L_{0}$), respectively. For $t \approx 1.5L_{0}$, a sudden jump and hysteresis appear due to changes in the number of magnetic solitons~\cite{CNS_1pich}, while no such features are observed for $t \approx L_{0}$.

Before detecting the spin phase with spin current, we evaluate the spin diffusion length in CrNb$_3$S$_6$ ($\lambda_{\rm CrNb_{3}S_{6}}$). We have performed NLSV measurements~\cite{NL0,NL1,NL2,NL3,NL4,Niimi_2015} with and without the CrNb$_3$S$_6$ thin flake, as illustrated in Figs.~2a and 2b. By applying an electric current $I$ from the left Py wire to the Cu wire, spin accumulation is generated at the interface between Py and Cu, which diffuses along the Cu wire, resulting in a spin current flow. When another Py wire is placed within the spin diffusion length, a nonlocal voltage ($V_{\rm NL1}$) dependent on the electrochemical potential is detected. When a CrNb$_3$S$_6$ flake is inserted as a middle wire (Fig.~2b), a part of the spin current is absorbed into the CrNb$_3$S$_6$ flake. As a result, the NLSV signal $\Delta R_{\rm S1} = \Delta V_{\rm NL1} / I$ is reduced compared to the signal without the CrNb$_3$S$_6$ flake because of the strong spin-orbit interaction of CrNb$_3$S$_6$, as shown in Fig.~2c.

The temperature dependence of $\Delta R_{\rm S1}$ for devices with and without the CrNb$_3$S$_6$ flake is shown in Extended Data Fig.~1. From the reduction rate, $\Delta R_{\rm S1}^{\rm with} / \Delta R_{\rm S1}^{\rm without}$, we have obtained $\lambda_{\rm CrNb_{3}S_{6}}$ using a one-dimensional spin diffusion model~\cite{Niimi_2015}. Figure~2d shows $\lambda_{\rm CrNb_{3}S_{6}}$ as a function of temperature. $\lambda_{\rm CrNb_{3}S_{6}}$ is approximately 5~nm at low temperatures. It decreases sharply above 50~K and is saturated at 1~nm above $T_{\rm C}$. The temperature dependence of the resistance $R$ of the CrNb$_3$S$_6$ thin flake is shown in the inset of Fig.~2d. $\lambda_{\rm CrNb_{3}S_{6}}$ is nearly inversely proportional to $R$, indicating that the Elliott-Yafet mechanism~\cite{Elliott,YAFET} is dominant in this material.

\section{detection of spin phase by spin current}

Next, we demonstrate spin phase detection using a lateral spin valve consisting of Py and CrNb$_3$S$_6$, as shown in Fig.~3a. In a conventional NLSV with two ferromagnetic wires, the NLSV signal changes the sign, depending on whether the injector and detector magnetizations are parallel or antiparallel. In our setup, we replace the detector Py wire by the chiral helimagnet CrNb$_3$S$_6$ flake. Since the spin current in CrNb$_3$S$_6$ decays within $\lambda_{\rm CrNb_{3}S_{6}}$ ($\approx$~5~nm) much smaller than the helical period $L_{0}$ ($=48$~nm) as discussed in the previous section, it is inherently surface-sensitive, enabling us to probe the surface magnetic moment direction and thus the spin phase.

Figure~3b shows the NLSV signal $R_{\rm S2} \equiv V_{\rm NL2} / I$ at low magnetic fields for the $t \approx L_{0}$ thick CrNb$_3$S$_6$ flake. A clear rectangular $R_{\rm S2}$ is observed. As $H_{y}$ increases from negative values, $R_{\rm S2}$ abruptly switches to negative at 50~Oe and returns to positive at 250~Oe (see the red curve). Similar behavior is observed during the reverse field sweep (see the blue curve). In the present device, the magnetization of the Py wire (left side in Fig.~1c) is known to flip at $H_{y} \approx 250$~Oe~\cite{NL2}, meaning that the jump at $H_{y} \approx \pm 50$~Oe originates from the CrNb$_3$S$_6$ flake. This jump in $R_{\rm S2}$ reflects a sign change in the $y$-component of the surface magnetic moment, indicating spin phase rotation by 180$\tcdegree$.

The 180$\tcdegree$ spin phase rotation under an external field suggests the existence of a net magnetization, indicating that a helical magnetic structure with a period shorter than $L_{0}$, rather than a perfect one-pitch helical structure, is realized. Magnetic pinning effects likely prevent forming a perfect one-pitch helix in such a thin flake. This interpretation is further supported by the absence of resistance jumps associated with changes in the number of solitons for the $t \approx L_{0}$ thick CrNb$_3$S$_6$ flake (see Fig.~1d).

Figure~3c shows the NLSV signal over a wider magnetic field range for another $t \approx L_{0}$ thick CrNb$_3$S$_6$ flake. Clear jumps in $R_{\rm S2}$ are also observed in this 52~nm thick sample, indicating that a helical structure with a period shorter than one pitch is realized even at this thickness. In addition, $R_{\rm S2}$ exhibits a gradual change from high to low magnetic fields, reflecting the continuous evolution of the surface magnetic moment direction in CrNb$_3$S$_6$. Surface barriers in chiral helimagnets can lead to a chiral surface twist~\cite{Shinozaki2018,chiraltwistcubic,PhysRevB.92.220412,CNS0}, causing this gradual rotation of surface magnetic moment direction at higher fields. The expected magnetic structures at characteristic magnetic fields during the magnetic field increasing process for the $t \approx L_{0}$ thick CrNb$_3$S$_6$ flake are illustrated in Fig.~3d. Additionally, the magnetoresistance of CrNb$_3$S$_6$ may influence $R_{\rm S2}$ through the spin resistance at the CrNb$_3$S$_6$/Cu interface~\cite{NL4}.

We then examine $R_{\rm S2}$ at low magnetic fields for a $t \approx 1.5L_{0}$ thick CrNb$_3$S$_6$ flake, shown in Fig.~3e. Similar to Fig.~3b, the jumps in $R_{\rm S2}$ at $\pm 250$~Oe originate from the magnetization flip of the Py wire. In the low magnetic field region, on the other hand, $R_{\rm S2}$ behaves differently from the case of $t \approx L_{0}$: it gradually decreases from $-50$~Oe (or $+50$~Oe), crosses zero near $0$~Oe, becomes negative, and flattens around $-100$~Oe (or $+100$~Oe) as $H_{y}$ increases (or decreases) further.
This gradual change in $R_{\rm S2}$ at low magnetic fields can be attributed to a continuous change of the $y$-component of the surface magnetic moment for the $1.5L_{0}$ helical magnetic structure. The $1.5L_{0}$ structure has the same net magnetization as a $0.5L_{0}$ helical structure. In the $0.5L_{0}$ structure, the surface magnetic moment is perpendicular to the net magnetization at $0$~Oe. This leads to $R_{\rm S2}=0$ at $\approx 0$~Oe. As the magnetic field increases, the surface magnetic moment rotates toward the field direction to increase the net magnetization. This rotation causes the gradual change in $R_{\rm S2}$.
The expected magnetic structures at characteristic low magnetic fields during the magnetic field increasing process for the $t \approx 1.5L_{0}$ thick CrNb$_3$S$_6$ flake are illustrated in Fig.~3g (see states 3-5). 

The NLSV signal over a wider magnetic field range for the $t \approx 1.5L_{0}$ thick CrNb$_3$S$_6$ flake is shown in Fig.~3f. When $H_{y}$ is increased from negative values (see the red data in Fig.~3e), $R_{\rm S2}$ exhibits jumps around $-500$~Oe and $800$~Oe, corresponding to magnetic fields where the number of magnetic solitons changes, as confirmed by the magnetoresistance shown in Fig.~1e. Additionally, $R_{\rm S2}$ gradually decreases from $-3000$~Oe to $-500$~Oe, likely due to the continuous evolution of the surface magnetic moment direction caused by the chiral surface twist (state 2 in Fig.~3g). After one magnetic soliton is inserted, the surface magnetic moment aligns parallel to the magnetic field, resulting in a positive $R_{\rm S2}$ once again (state 3 in Fig.~3g). A similar behavior due to the chiral surface twist is also observed after soliton extraction (state 6 in Fig.~3g). Changes in the resistance $R$ of CrNb$_3$S$_6$ may also contribute to the variations in $R_{\rm S2}$. A similar behavior is observed during the opposite field sweep (see the blue data in Fig.~3e).

As discussed in this section, the observed magnetic field dependence of $R_{\rm S2}$ for the $t \approx L_{0}$ and $t \approx 1.5L_{0}$ thick CrNb$_3$S$_6$ flakes is consistent with the expected behavior of the surface magnetic moment in a chiral helimagnet. These results confirm the successful spin phase detection using spin currents in NLSV devices. This is further corroborated by micromagnetic simulations in the next section. The reproducibility of these findings has been confirmed in different devices, as shown in Extended Data Fig.~2.

\section{Micromagnetic Simulations}

To further verify the magnetic field dependence of $R_{\rm S2}$, we have calculated magnetic structures for different helical magnetic structure lengths $L_z$. Details of the simulations are provided in the METHODS section. Since the experimental results indicate that a helical magnetic structure with a period shorter than $L_{0}$, rather than a perfect one-pitch helical structure, is realized in $t \approx L_{0}$ thick CrNb$_3$S$_6$ flakes, we have performed simulations for the $L_z = 7\,(\approx 0.5L_{0})$ helical magnetic structure.

Figures~4a shows snapshots of the magnetization components along the $x$- and $y$-directions ($M_{x}$ and $M_{y}$, respectively) and magnetic structure averaged over 10 spins at characteristic magnetic fields during the field-increasing process. At $H = -H_0$, which is large enough to prepare a forced ferromagnetic state, all spins align nearly along the magnetic field direction ($-y$ direction)  (state 1). As the magnetic field increases, the surface magnetic moments at the edges of the $L_z$ direction begin canting toward the $x$-axis due to the chiral surface twist (state 2). At a small positive magnetic field, a 180$\tcdegree$ spin phase rotation occurs (state 3). After this spin phase switching, the surface magnetic moments gradually realign with the magnetic field (state 4).

These behaviors are consistent with the expected magnetic structure for the $t \approx L_{0}$ thick CrNb$_3$S$_6$ flake inferred from the experimental $R_{\rm S2}$ data, as illustrated in Fig.~3d. This confirms that spin currents in NLSV devices can effectively detect the surface magnetic moment, i.e., the spin phase.

To justify the use of $L_z = 7\,(\approx 0.5L_{0})$ in our simulations, we have also performed calculations for the $L_z = 14\,(\approx L_{0})$ helical magnetic structure, as shown in Extended Data Fig.~3. In this case, the surface magnetic moments rotate toward the $+y$ direction before reaching 0~Oe during the increasing field process (see state2 in Extended Data Fig.~3). As a result, the system forms a complete one-pitch helical structure at 0~Oe. Moreover, no 180$\tcdegree$ spin phase rotation is observed. This behavior contrasts with the experimental results, supporting our interpretation that the realized magnetic structure has a period shorter than $L_{0}$.

Next, we investigate the $t \approx 1.5L_{0}$ thick CrNb$_3$S$_6$ flake by analyzing the $L_z = 21\,(\approx 1.5L_{0})$ helical magnetic structure, as shown in Fig.~4b. As the magnetic field increases from $H = -H_0$ (state 1), the surface magnetic moments rotate toward the $+y$ direction before soliton insertion due to the chiral surface twist (state 2). After the soliton insertion, the surface magnetic moments align along the magnetic field to increase the net magnetization. As a result, they gradually rotate from the $-y$ to the $+y$ direction (states 3-5). The surface twisted configuration immediately after soliton extraction, shown as state 6 in Fig.~3g, does not appear in our simulations. This is because the soliton still remains even at higher magnetic fields (state 6$'$ in Fig.~4b). The larger soliton extraction field in the simulation compared to the experiment could be due to differences in physical parameters such as temperature and the number of spins between the two cases~\cite{Mito2018}.

These behaviors, except for state 6, closely resemble the expected magnetic structures for the $t \approx 1.5L_{0}$ thick CrNb$_3$S$_6$ flake, as inferred from the experimental $R_{\rm S2}$ data and illustrated in Fig.~3g. This further confirms spin currents effectively detect the surface moment, i.e., the spin phase.

\section{Spin Hall Effect}

To investigate spin fluctuations in chiral helimagnets~\cite{CNS2w}, we have measured the ISHE in CrNb$_3$S$_6$ flakes using the spin absorption method, as illustrated in Fig.~5a. By applying a current $I$ from the left Py wire to the Cu wire, part of the spin current is injected into CrNb$_3$S$_6$. When the Py magnetization is fully polarized along the Cu wire ($|H_x| > 3000$~Oe), spin-to-charge conversion via the ISHE occurs in CrNb$_3$S$_6$.

Figure~5b shows the ISHE signals of the 72~nm-thick CrNb$_3$S$_6$ device measured at $T = 130$~K and 300~K. The ISHE resistance, $R_{\rm ISHE}$, is defined as the detected voltage drop $V_{\rm ISHE}$ along the CrNb$_3$S$_6$ flake divided by $I$. When $|H_x| > 3000$~Oe, $R_{\rm ISHE}$ saturates. At room temperature, a clear negative ISHE signal $[2\Delta R_{\rm ISHE} \equiv R_{\rm ISHE} (H_{x} > 3000~{\rm Oe}) - R_{\rm ISHE} (H_{x} < -3000~{\rm Oe})]$ is observed, consistent with the behavior of pure Nb~\cite{NL3}. However, near the critical temperature $T_{\rm C} = 130$~K, the ISHE signal reverses its sign, becoming positive.

Figure~5c shows the temperature dependence of $\Delta R_{\rm ISHE}$ for CrNb$_3$S$_6$ devices with two different thicknesses. We also estimated the spin Hall angle, $\alpha_{\rm H}$, assuming an interface shunting factor of approximately 0.2 between Cu and CrNb$_3$S$_6$~\cite{NL3}, as shown in Fig.~5d.
For both devices, $\Delta R_{\rm ISHE}$ and $\alpha_{\rm H}$ peak at $T_{\rm C} = 130$~K. The sign of $\Delta R_{\rm ISHE}$ and $\alpha_{\rm H}$ remains negative for the 43~nm-thick CrNb$_3$S$_6$ device, whereas a sign reversal occurs near $T_{\rm C}$ for the 72~nm-thick device. Reproducibility across other devices is demonstrated in Extended Data Fig.~4. These results suggest that magnetic fluctuations near $T_{\rm C}$ lead to the ISHE sign reversal in the thicker device. 

Since a full helical pitch isn’t realized in the 43nm flake, as discussed in previous sections, the observed magnetic fluctuations may be associated with helical magnetic ordering. Similar behavior around the critical temperature has been reported in ferromagnets~\cite{PdNi,FePt,NiCu} and antiferromagnets~\cite{SHE_Cr}, where magnetic fluctuations play a significant role~\cite{SHE_Nagaosa}. While magnetic fluctuations are also expected in our case, further studies are needed to elucidate their precise role.

Before closing this section, we address the potential influence of thermoelectric signals in ISHE measurements with a lateral spin valve structure, particularly in magnetic materials~\cite{Co2MnGa}. Heat currents at the spin injection interface can generate voltages at the detection interface through the anomalous Nernst effect in addition to the ISHE. However, the temperature dependence of $\Delta R_{\rm ISHE}$ in our measurements does not follow that of the CrNb$_3$S$_6$ magnetization~\cite{CNS4}, excluding thermoelectric effects as a dominant contribution to $R_{\rm ISHE}$.

\section{conclusion}
In conclusion, we have demonstrated the spin phase detection in a vdW nanoscale chiral helimagnet CrNb$_3$S$_6$ using NLSV devices. Due to its short spin diffusion length ($\approx$~5~nm), we successfully detected the surface magnetic moment direction, corresponding to the spin phase.
NLSV measurements on CrNb$_3$S$_6$ flakes with thicknesses of nearly $L_{0}$ and $1.5L_{0}$ have revealed a non-monotonic magnetic field dependence of spin signal, consistent with the field-dependent behavior of the surface magnetic moment in $0.5L_{0}$ and $1.5L_{0}$ helical magnetic structures, as supported by micromagnetic simulations. Furthermore, ISHE measurements have detected a sign change in the spin Hall angle near $T_{\rm C}$, attributed to spin fluctuations in the chiral helimagnet.
These results establish spin currents as a powerful probe for spin phase and fluctuations, demonstrating their nanoscale effectiveness. Moreover, our findings highlight the spin phase as a key internal degree of freedom in nanoscale spintronic devices, paving the way for new research directions in spintronics.

%

\newpage
\section{methods}
\subsection{Device fabrication}
CrNb$_3$S$_6$ single crystals were grown by the chemical transport method, as described elsewhere~\cite{CNS6}. CrNb$_3$S$_6$ thin flakes were obtained by the conventional mechanical exfoliation method using Nitto tapes. These thin flakes were transferred onto a SiO$_2$ (285~nm)/Si substrate. The exfoliation process was performed inside a glove box filled with Ar gas of 99.9999\% purity to prevent oxidation of the CrNb$_3$S$_6$ thin flakes.

The spin valve structure was fabricated using a conventional lift-off method. Two Py wires and a Cu bridge were patterned by electron beam lithography onto the substrate coated with polymethyl-methacrylate (PMMA) resist for the deposition of Py and Cu. A pair of Py wires was first deposited on both sides of the CrNb$_3$S$_6$ thin flake using an electron beam evaporator. The width and thickness of the Py wires are 100 and 30~nm, respectively. Before the deposition of the Cu bridge, Ar ion beam milling was performed for approximately 1 minute to remove the residual resist and clean the surfaces of the Py wires and CrNb$_3$S$_6$ thin flakes. After the Ar ion beam milling, the device was transferred to another chamber without breaking vacuum, and the Cu bridge was deposited by a Joule heating evaporator. The width of the Cu bridge is 100~nm, and the thickness is either 100 or 200~nm. The width of the CrNb$_3$S$_6$ thin flakes is approximately 400~nm, and the thickness $t$ of the CrNb$_3$S$_6$ thin flakes used in this work is 43~nm, 50~nm, 52~nm, 72~nm and 73~nm (see Extended Data Table 1). We also fabricated lateral spin valve devices without CrNb$_3$S$_6$ flakes as reference devices to evaluate the spin diffusion length. The device structure, such as the width and thickness of the Py and Cu wires, is the same as that of the devices with CrNb$_3$S$_6$.

\subsection{Measurement set-up}

The nonlocal spin valve and inverse spin Hall effect measurements were conducted using an ac lock-in amplifier and a $^4$He flow cryostat. A magnetic field in the range of $\pm 0.4$~T was applied to the device using an electromagnet.

\subsection{Details of micromagnetic simulations}

To verify the magnetic field dependence of $R_{\rm S2}$, we performed numerical simulations of the magnetic structure of a chiral helimagnet under an applied magnetic field. We employed the classical Heisenberg spin system with the following Hamiltonian:
\begin{equation}
\mathcal{H} = - \sum_{\langle i,j \rangle} J \vec{S}_i \cdot \vec{S}_j 
+ \sum_{\langle i,j' \rangle} \vec{D} \cdot (\vec{S}_i \times \vec{S}_{j'})
+ \sum_{i} K_z S_{iz}^2,
\end{equation}
where $\vec{S}_i$ is the spin of the $i$ site.
$J$ and $\vec{D}=(0,0,D)$ represent the ferromagnetic exchange interaction and the DM interaction.
We use the 2-dimensional spin system consists of $L_z \times 10$ spins, where $L_z$ is the system length corresponds to thickness in the experiments.
$\langle i,j \rangle$ takes nearest neighbor sites whereas $\langle i,j' \rangle$ takes nearest neighbor sites only in the $z$ direction.
$K_{z}$ is the anisotropy energy along the hard $z$-axis, which stabilizes the helical spin structure with a propagation vector along the $z$-axis.

Using the Landau-Lifshitz-Gilbert equation, we numerically calculated the magnetic structure under a magnetic field applied parallel to the $y$-axis (parallel to the helical plane). The values of $J$ and $D$ were both set to $6.2$\,meV resulting in a helical magnetic structure with a period $(L_0)$ of approximately 13.3 spins in the 2-dimensional system. Additionally, thermal fluctuations of 10~K were incorporated by introducing a random magnetic field~\cite{PhysRevB.58.14937}.
We set the $K_z=6.2 \times 10^{-2}$\,meV that is enough small compared with $J$  and $D$.

\section{data availability}
Data are available from the corresponding authors upon request.

\section{Acknowledgments}
The authors thank J. Kishine, S. Maekawa, and S. Okumura for fruitful discussions. The crystal structure of CrNb$_3$S$_6$ was visualized using VESTA~\cite{vesta}. This work was supported by JSPS KAKENHI (Grant Nos. JP22H04481, JP23H00257, JP23K13062, JP23H00091, JP23H01870), JST FOREST (Grant No. JPMJFR2134), and the Cooperative Research Project of RIEC, Tohoku University.

\section{Author contributions}
N.J., S.S., I.S., K.Y., and R.K. performed the device fabrication and measurements. Y.S., H.S., Y.K., and Y.T. synthesized the CrNb$_3$S$_6$ cystals. J.O. conducted the micromagnetic simulations. N.J. wrote the manuscript with input from S.T., J.O., Y.T., and Y.N.

\clearpage
\newpage

\onecolumngrid

\begin{figure}
\includegraphics[width=170mm]{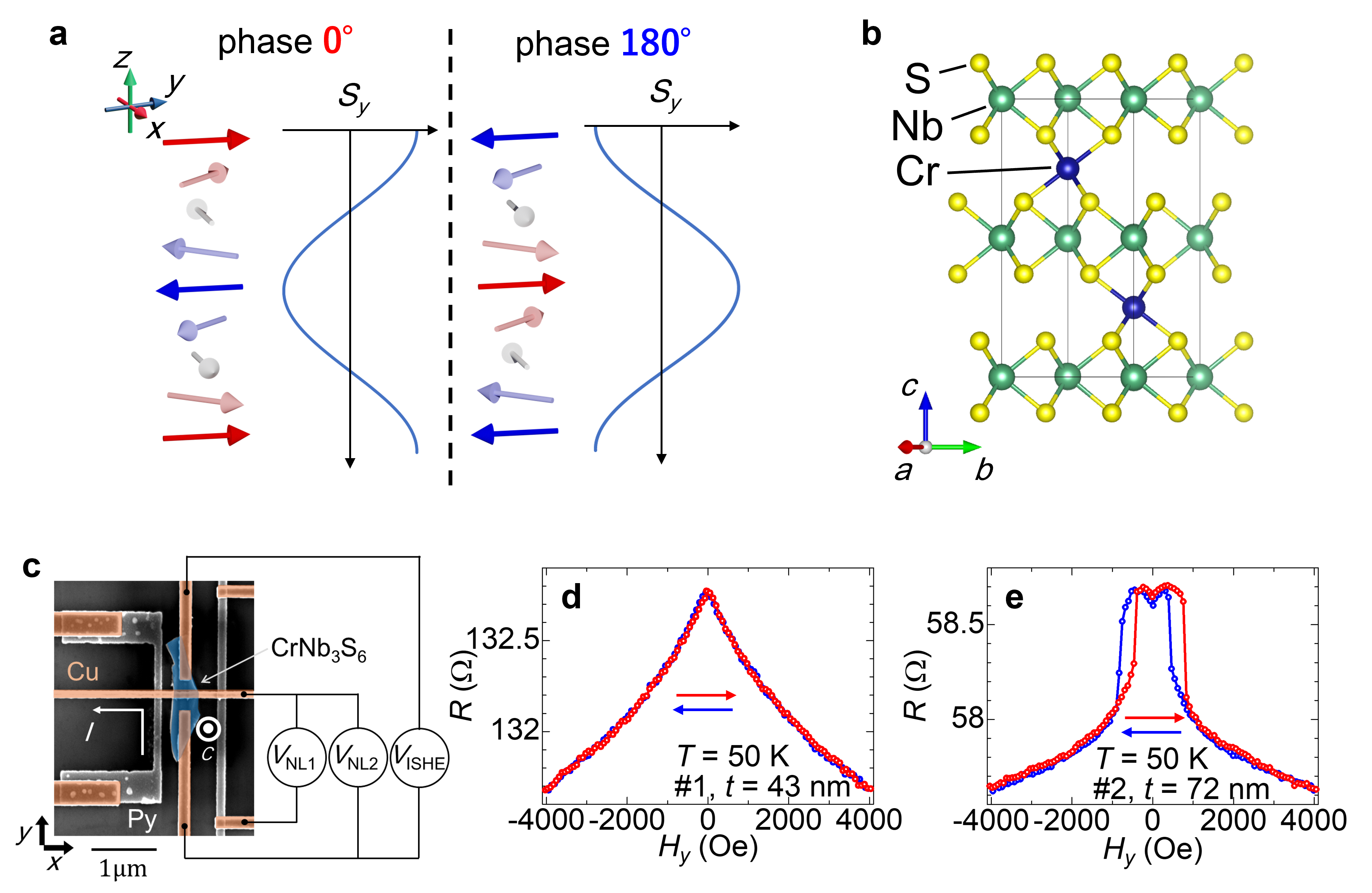}
\caption{\textbf{Basic properties of CrNb$_3$S$_6$.} 
\textbf{a}, Definition of the spin phase in a helical magnetic structure. The 0$\tcdegree$ and 180$\tcdegree$ phases are defined when the surface magnetic moment is parallel and antiparallel to an in-plane reference axis (taken here as the $y$-axis), respectively. \textbf{b}, Crystal structure of CrNb$_3$S$_6$. The black solid line represents a unit cell. \textbf{c}, A scanning electron microscope image of a typical nonlocal spin valve (NLSV) device. The scale bar is 1~$\mu$m. \textbf{d, e}, Magnetoresistance of CrNb$_3$S$_6$ with thicknesses of 43~nm (\textbf{d}) and 72~nm (\textbf{e}) measured at 50~K. The red and blue data correspond to the magnetic field increasing and decreasing processes, respectively.
}
\end{figure}

\begin{figure}
\includegraphics[width=150mm]{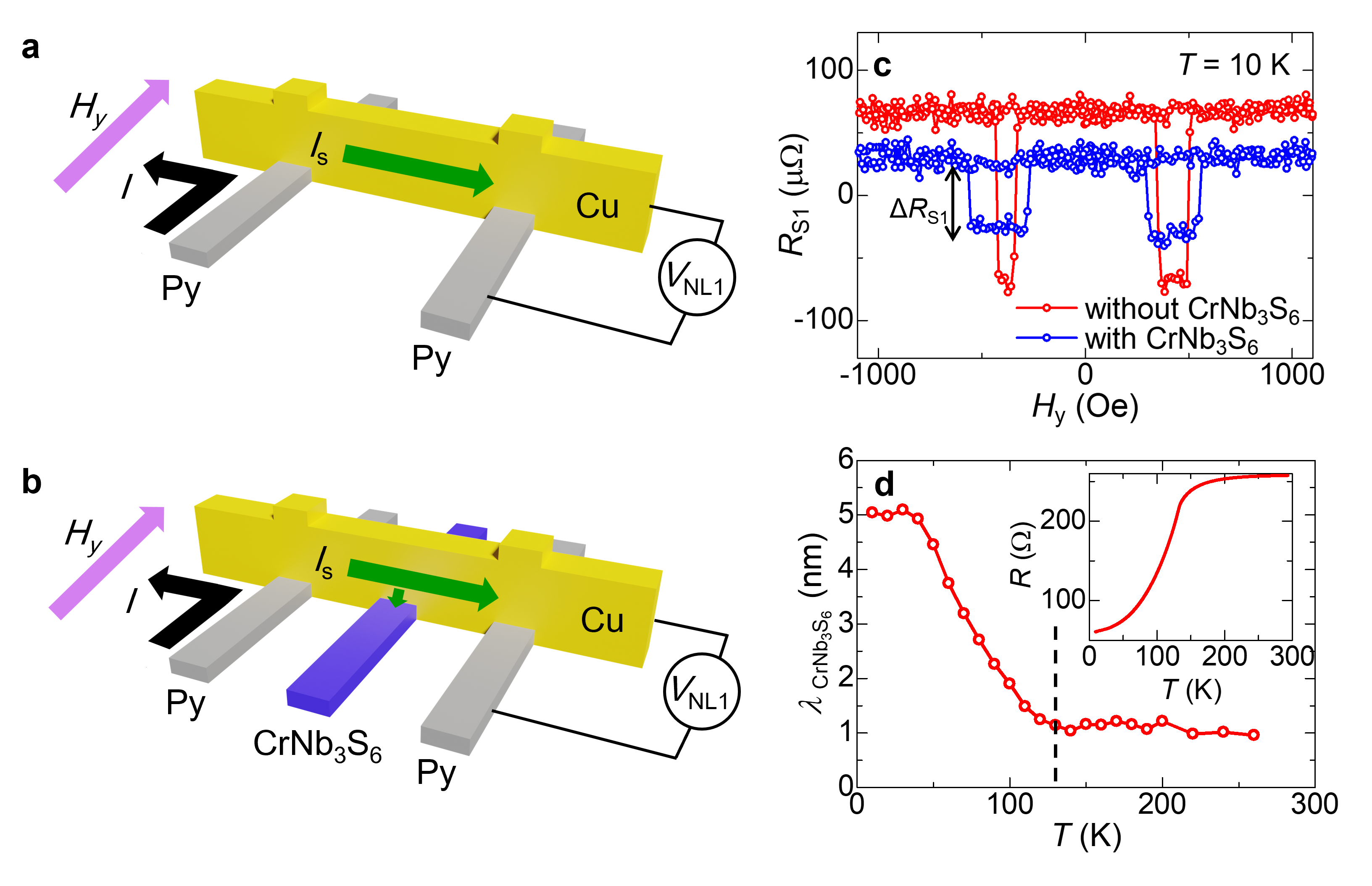}
\caption{\textbf{Determination of spin diffusion length of CrNb$_3$S$_6$.} 
\textbf{a, b}, Schematic illustrations of NLSV measurements without (\textbf{a}) and with (\textbf{b}) an insertion of a CrNb$_3$S$_6$ flake. \textbf{c}, NLSV signal $R_{\rm S1}$ with (blue) and without (red) a 30~nm thick CrNb$_3$S$_6$ flake measured at $T = 10$~K. \textbf{d}, Spin diffusion length of the CrNb$_3$S$_6$ flake as a function of temperature. The inset shows the temperature dependence of the resistance $R$ of the CrNb$_3$S$_6$ flake.}
\end{figure}

\begin{figure}
\includegraphics[width=150mm]{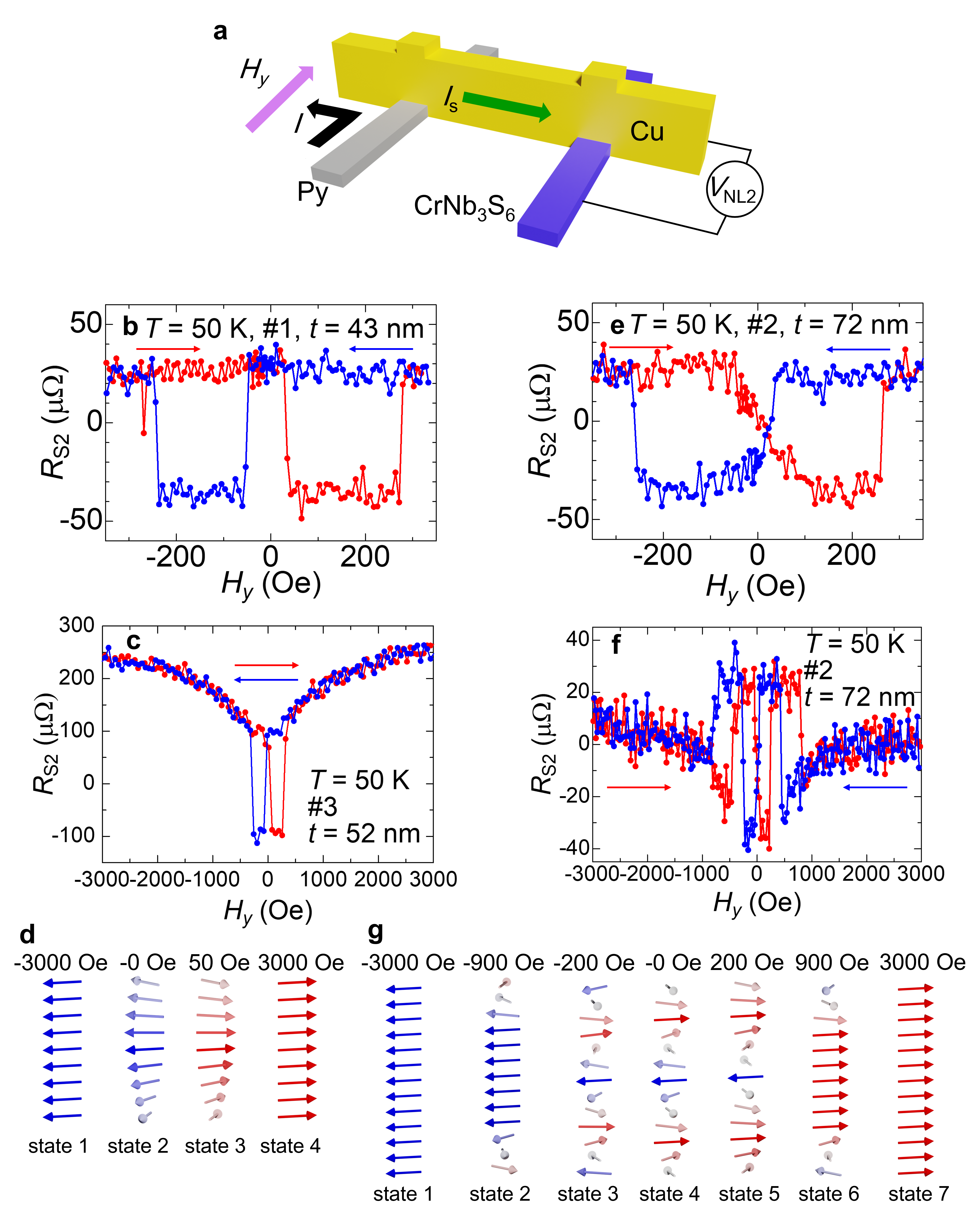}
\caption{\textbf{Spin phase detection in CrNb$_3$S$_6$ flakes.} 
\textbf{a}, A schematic illustration of the spin phase detection measurement. \textbf{b, c}, NLSV signal $R_{\rm S2}$ at $T = 50$~K using a 43~nm ($\approx L_{0}$) thick CrNb$_3$S$_6$ flake for a low magnetic field region (\textbf{b}) and a 52~nm ($\approx L_{0}$) thick CrNb$_3$S$_6$ flake for a high magnetic field region (\textbf{c}). \textbf{d}, The expected magnetic structures at characteristic magnetic fields during the magnetic field increasing process for the $t \approx L_{0}$ thick CrNb$_3$S$_6$ flake. \textbf{e, f}, NLSV signal $R_{\rm S2}$ at $T = 50$~K using a 72~nm ($\approx 1.5L_{0}$) thick CrNb$_3$S$_6$ flake for a low magnetic field region (\textbf{e}) and a high magnetic field region (\textbf{f}).
\textbf{g}, The expected magnetic structures at characteristic magnetic fields during the magnetic field increasing process for the $t \approx 1.5L_{0}$ thick CrNb$_3$S$_6$ flake. 
}
\end{figure}

\begin{figure}
\includegraphics[width=180mm]{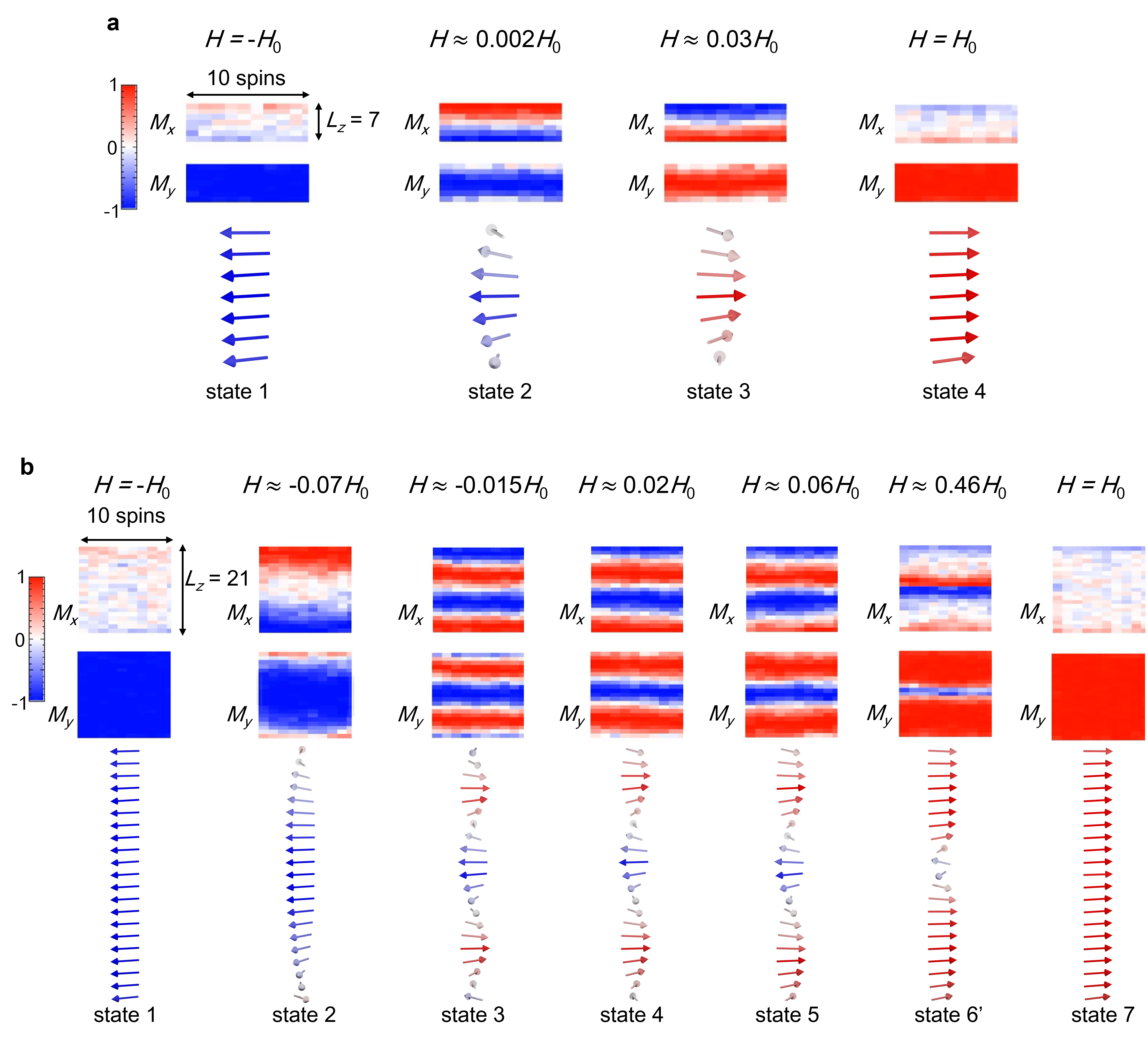}
\caption{\textbf{Micromagnetic simulations.} 
\textbf{a}, Snapshots of the magnetization components along the $x$- and $y$-directions ($M_{x}$ and $M_{y}$, respectively) and magnetic structure averaged over 10 spins at characteristic magnetic fields during the field-increasing process for a $L_z = 7\,(\approx 0.5L_{0})$ helimagnet. The maximum values of $M_{x}$ and $M_{y}$ are normalized to 1.
\textbf{b}, Snapshots of $M_{x}$ and $M_{y}$ and magnetic structure averaged over 10 spins at characteristic magnetic fields during the field-increasing process for a $L_z = 21\,(\approx 1.5L_{0})$ helimagnet. The maximum values of $M_{x}$ and $M_{y}$ are normalized to 1.
}
\end{figure}

\begin{figure}
\includegraphics[width=150mm]{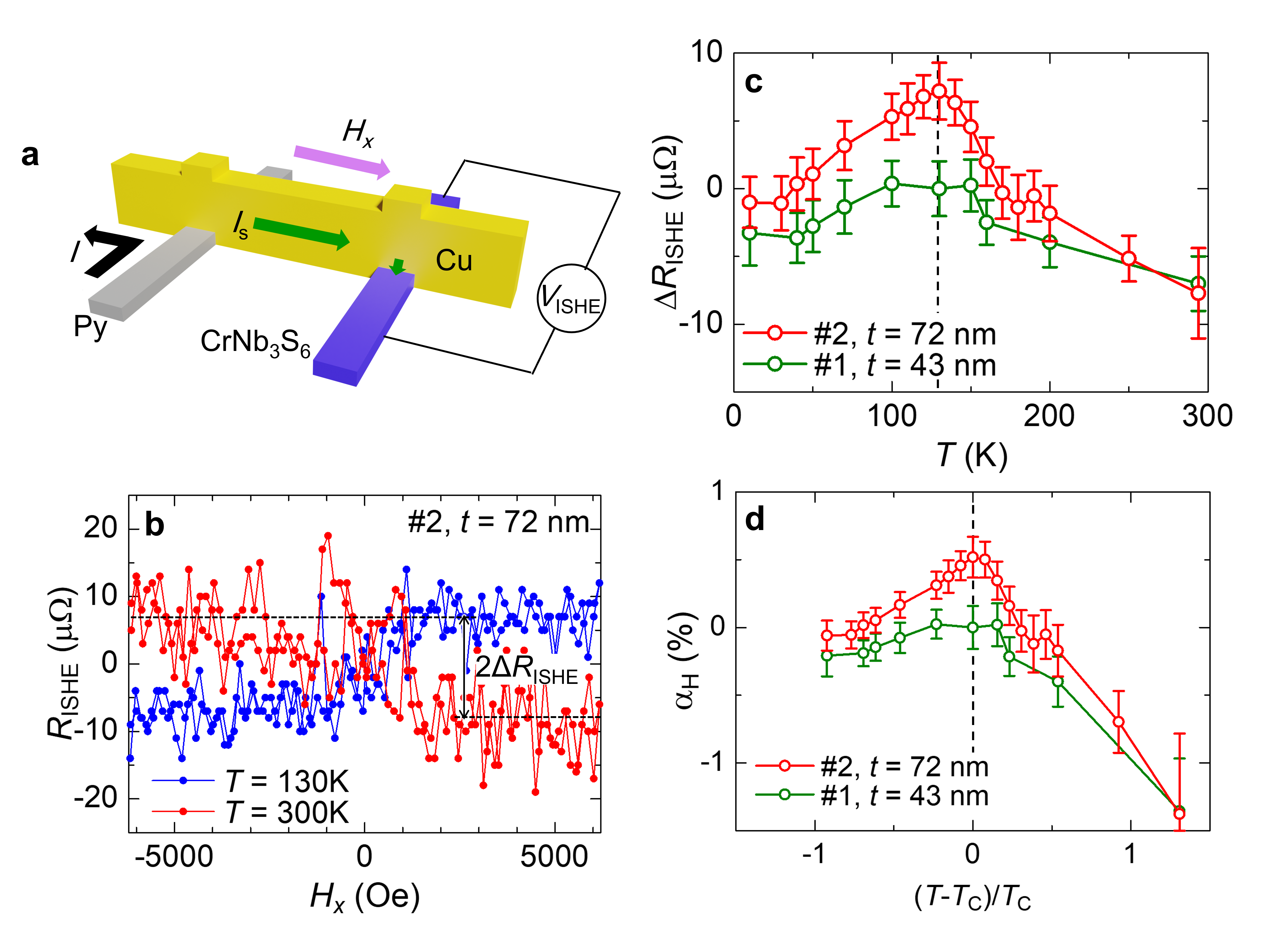}
\caption{\textbf{Inverse spin Hall effect measurements in CrNb$_3$S$_6$ flakes.} 
\textbf{a}, A schematic illustration of the ISHE measurement. \textbf{b}, $R_{\rm ISHE}$ of a 72~nm-thick CrNb$_3$S$_6$ device measured at $T = 130$ and 300~K. $\Delta R_{\rm ISHE}$ is defined in the figure. \textbf{c}, Temperature dependence of $\Delta R_{\rm ISHE}$ for $t = 43$ and 72~nm thick CrNb$_3$S$_6$ devices. \textbf{d}, The spin Hall angle $\alpha_{\rm H}$ as a function of the reduced temperature, $(T - T_{\rm C})/T_{\rm C}$, for $t = 43$ and 72~nm thick CrNb$_3$S$_6$ devices.}
\end{figure}

\twocolumngrid

\clearpage
\newpage

\onecolumngrid
\clearpage
\newpage
\section{Extended Data}

\captionsetup[table]{name=Extended Data Table}
\setcounter{table}{0}

\begin{table}[h]
\centering
\begin{tabular}{cccccc}
\toprule
Sample No. & Thickness (nm) & Spin Signal & Spin Hall \\
\midrule
1 & 43 ($\approx L_{0}$) & \checkmark & \checkmark \\
2 & 72 ($\approx 1.5L_{0}$) & \checkmark & \checkmark \\
3 & 52 ($\approx L_{0}$) & \checkmark & \checkmark \\
4 & 50 ($\approx L_{0}$) & \checkmark & \checkmark \\
5 & 50 ($\approx L_{0}$) & \checkmark &   \\
6 & 73 ($\approx 1.5L_{0}$) & \checkmark & \checkmark \\
\bottomrule
\end{tabular}
\caption{Thickness of the measured samples and the corresponding measurements performed.
}
\end{table}

\captionsetup[figure]{name=Extended Data Fig.}
\setcounter{figure}{0}

\begin{figure}[H]
\centering
\includegraphics[width=85mm]{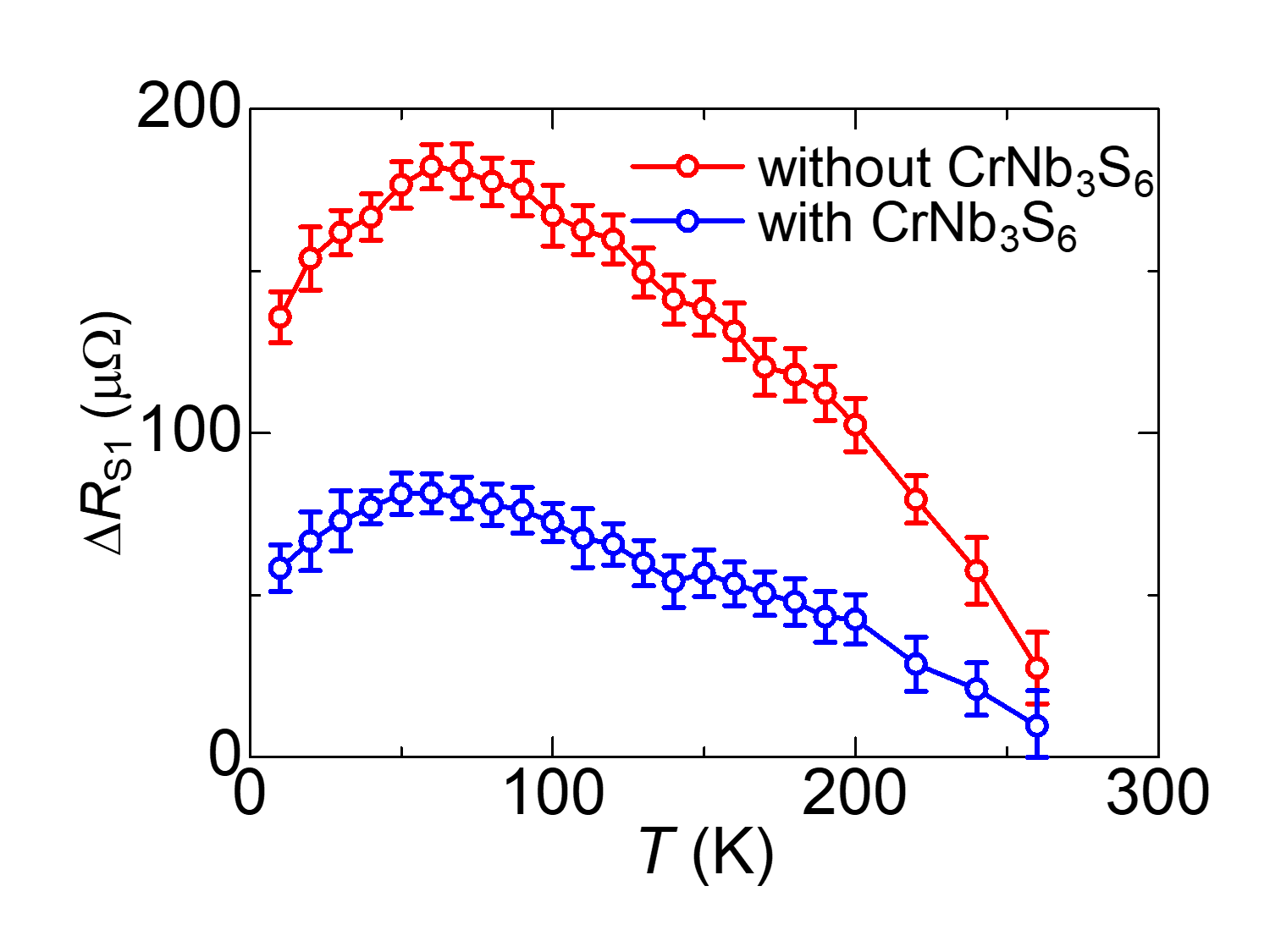}
\extendedcaption{\textbf{Temperature dependence of the NLSV signal with and without a 30~nm thick CrNb$_3$S$_6$ flake.} \label{fig:epsart} The temperature dependence of $\Delta R_{\rm S1}$ for devices with and without the CrNb$_3$S$_6$ flake.}
\end{figure}

\begin{figure}[H]
\centering
\includegraphics[width=80mm]{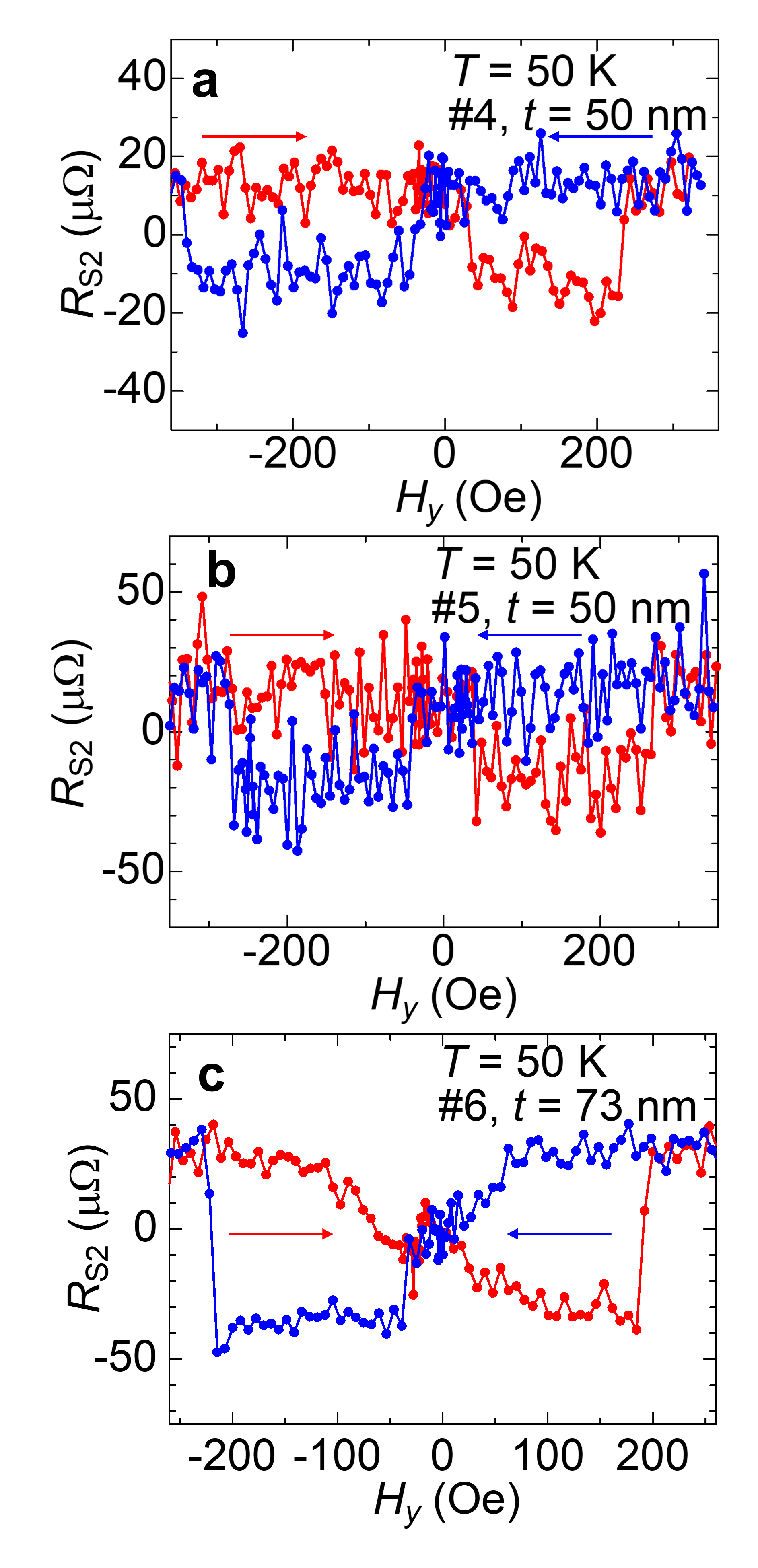}
\extendedcaption{\textbf{Reproducibility of the NLSV signal using CrNb$_3$S$_6$ flakes.} \label{fig:epsart} NLSV signals $R_{\rm S2}$ at $T = 50$~K using $t\approx L_{0}$ (\textbf{a, b}) and $t\approx 1.5L_{0}$ (\textbf{c}) CrNb$_3$S$_6$ flakes.
}
\end{figure}

\begin{figure}[H]
\centering
\includegraphics[width=150mm]{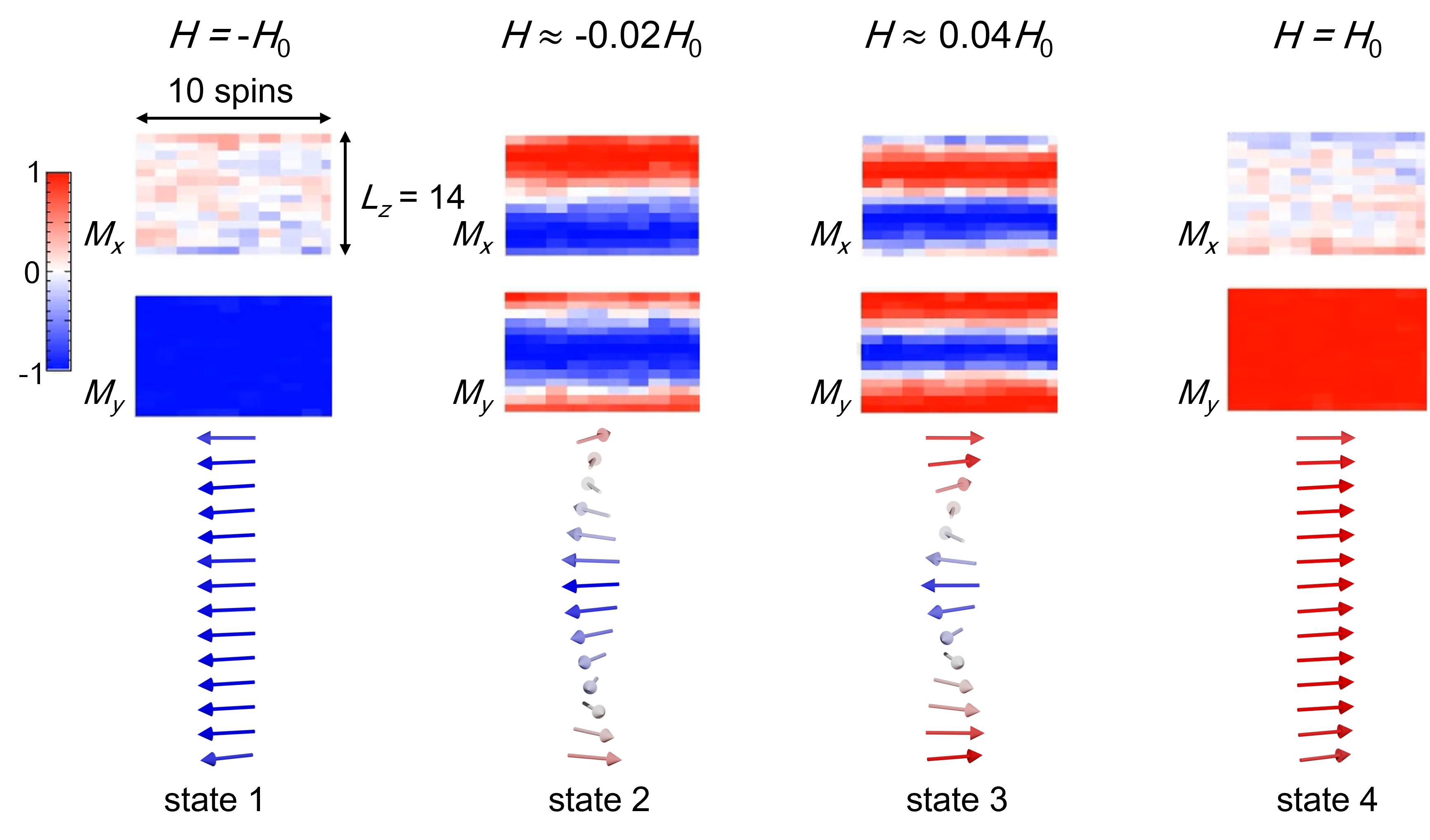}
\extendedcaption{\textbf{Micromagnetic simulations for a $L_z = 14\,(\approx L_{0})$ helimagnet.} \label{fig:epsart}  Snapshots of the magnetization components along the $x$- and $y$-directions ($M_{x}$ and $M_{y}$, respectively) and magnetic structure averaged over 10 spins at characteristic magnetic fields during the field-increasing process for a $L_z = 14\,(\approx L_{0})$ helimagnet. The maximum values of $M_{x}$ and $M_{y}$ are normalized to 1.
}
\end{figure}

\begin{figure}[H]
\centering
\includegraphics[width=150mm]{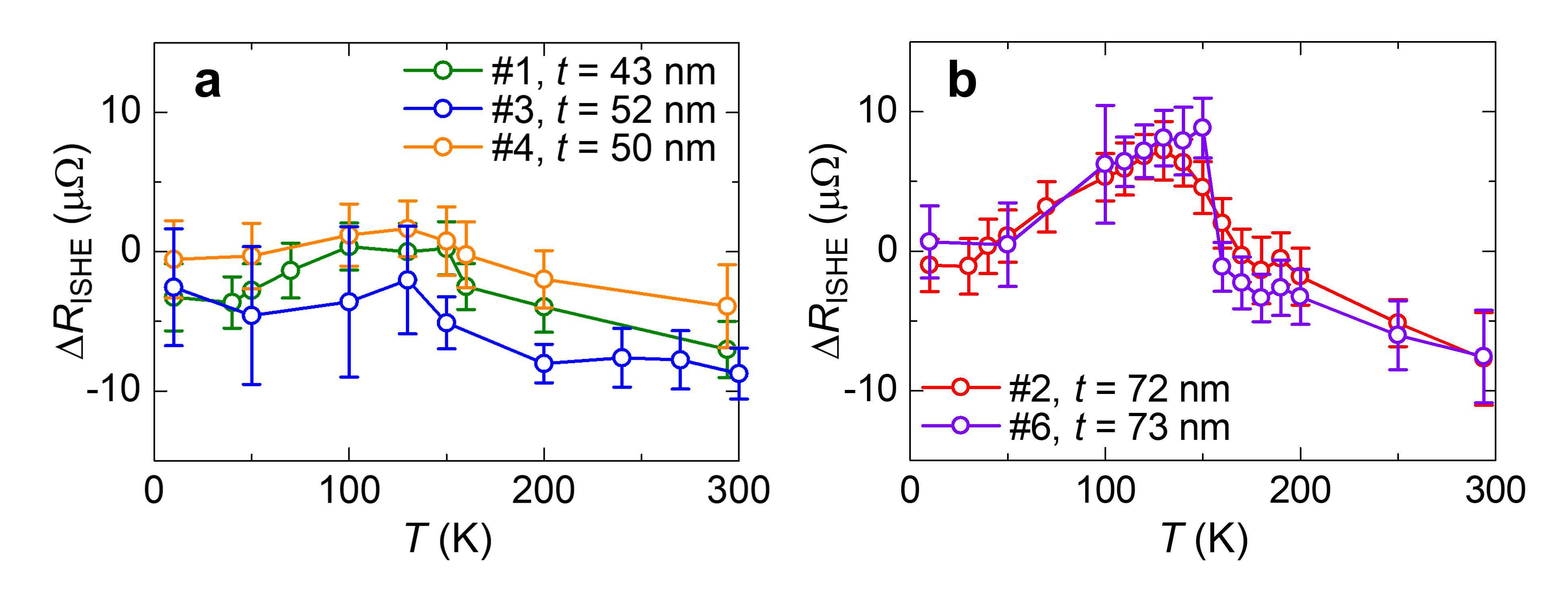}
\extendedcaption{\textbf{Reproducibility of inverse spin Hall effect.} \label{fig:epsart} Temperature dependence of $\Delta R_{\rm ISHE}$ for $t\approx L_{0}$ (\textbf{a}) and $t\approx 1.5L_{0}$ (\textbf{b}) CrNb$_3$S$_6$ devices. The data for sample \#1 and \#2 are the same as those shown in the main text.}
\end{figure}

\end{document}